\documentclass[aps,prd,english,superscriptaddress, longbibliography,11pt,notitlepage, nofootinbib]{revtex4}

	\usepackage[colorlinks=true, a4paper=true, pdfstartview=FitV,
linkcolor=blue, citecolor=blue, urlcolor=blue]{hyperref}

\usepackage{amsmath}
\usepackage{amssymb}
\usepackage{graphicx}
\usepackage{hhline}
\usepackage{mwe}
\usepackage{amsbsy}
\usepackage{textcomp}
\usepackage{commath}
\usepackage{subfig}
\usepackage{slashed}
\usepackage{natbib}

\usepackage{stackengine}
\stackMath

\makeatletter
\usepackage{babel}
\numberwithin{equation}{section}
\begin{document}
\immediate\write16{<<WARNING: LINEDRAW macros work with emTeX-dvivers
                    and other drivers supporting emTeX \special's
                    (dviscr, dvihplj, dvidot, dvips, dviwin, etc.) >>}

\title{ Collective coordinates method for long-range kink collisions}
\author{J. G. F. Campos}
\email{joao.gfcampos@upe.br}
\affiliation{Física de Materiais, Universidade de Pernambuco, Rua Benfica, 455, Recife - PE - 50720-001, Brazil}
\author{A. Mohammadi}
\email{azadeh.mohammadi@ufpe.br}
\affiliation{Departamento de Física, Universidade Federal da Pernambuco, Av. Prof. Moraes Rego, 1235, Recife - PE - 50670-901, Brazil}
\author{T. Romanczukiewicz}
\email{tomasz.romanczukiewicz@uj.edu.pl}
\affiliation{Jagiellonian University, Krakow, Poland}

\begin{abstract}

In this paper, we explored a class of potentials with three minima that support kink solutions exhibiting one long-range tail. We analyzed antikink-kink interactions using an effective Lagrangian based on collective coordinates and compared the results to those obtained from full dynamical simulations. To this end, we constructed the collective coordinates with the antikink-kink configuration, and also a generalized Derrick mode, choosing the kink position and the Derrick mode amplitude as the moduli. For the antikink-kink configuration, we utilized the impurity ansatz proposed in \cite{campos2024collision}. We also studied the interaction of wobbling kinks where the lowest delocalized mode is excited. 

\end{abstract}

\maketitle

\section{Introduction}
\label{intro}

Kinks, topological solitons in $1+1$ dimensions, have been the subject of intensive research for over forty years due to their rich mathematical properties and complex interactions \cite{rajaraman1982solitons,manton2004topological,vachaspati2007kinks,shnir2018topological,kevrekidis2019four}. In integrable cases, where sufficient symmetries are present, kink interactions are relatively predictable, often resulting in a simple crossing with no loss as radiation. On the other hand, a large variety of phenomena can appear when the interaction ingredients are non-integrable. In this case, the system's dynamics can become chaotic. For instance, in kink-antikink collisions, outcomes vary with initial velocity, resulting in either reflection, annihilation through bion formation, or direct conversion into radiation. The former happens above a critical velocity, and the latter for a small initial velocity. A middle region can form, alternating between these two possibilities with chaotic structure and even fractal dimension. This region exists when there can be an interplay between translational and vibrational energy. The energy exchange mechanism can have several sources, including the shape modes of each kink in the interaction, as first explained in the seminal papers \cite{Sugiyama,campbell1983resonance,peyrard1983kink,campbell1986kink}, the delocalized bound modes of the pair \cite{dorey2011kink},  quasinormal modes \cite{dorey2018resonant,campos2020quasinormal},  sphalerons \cite{adam2021sphalerons} or even fermion fields at excited bound states \cite{bazeia2022resonance}. 

One approach to simplifying topological soliton dynamics is to use a collective coordinate model, also known as moduli space dynamics. In this method, instead of dealing with an infinite number of field degrees of freedom in the original field theory Lagrangian, the system is reduced to a finite number $N$ of key parameters, or ``moduli''. This reduction is effective when only a select set of field configurations contributes significantly to the soliton's behavior, allowing the rest to be ignored. In this method, the moduli become time-dependent, and by substituting these configurations into the Lagrangian and integrating it over space, one obtains an effective Lagrangian for motion in the $N$-dimensional moduli space. This creates a simplified, finite-dimensional mechanical model for a particle in a curved space and a possible interaction potential that can describe the dynamics of interacting solitons. The first successful approach for the kinks was presented in \cite{manton2021collective}, qualitatively describing symmetric kink-antikink scattering in the $\phi^4$ model. In this approximation method, two important issues needed to be addressed: the known null vector problem and the appropriate initial conditions \cite{manton2021kink}. 
In \cite{adam2022relativistic}, the authors explored the relativistic
collective coordinate model for multi-kink dynamics, including a Derrick mode in the set of moduli space configurations. This way, they could capture the Lorentz contraction factor needed for the dynamics. The method was successfully applied to the $\phi^6$ model where the resonance structure appears thanks to including the lowest delocalized mode \cite{adam2022multikink}. More recently, the authors could improve the results with perturbative relativistic moduli space, also promoting the amplitudes of the higher-order Derrick modes to collective coordinates \cite{adam2023relativistic}. In this line of work, novel results appeared considering a moduli space with a boundary \cite{adam2023moduli} and collisions with nonzero total momentum \cite{adam2023moduli2}.

Kinks can be short-range, where the tails tend to the vacua exponentially, or long-range, with one or two power-law tails. The main difficulty of most long-range kink models is that there is no explicit analytical form for the kinks. In the kink-antikink interactions, the force is exponentially suppressed with the separation distance in the former case and power-law in the latter one \cite{manton2019forces,d2020forces,manton2024antikink}. Long-range kinks and their interactions have attracted significant attention due to their complexity and unique phenomena associated with them \cite{braun1990kinks,woafo1993kink,mello1998topological, bazeia2023kink,christov2019long,campos2021interaction,christov2021kink,khare2022kink,bazeia2023geometrically,campos2024collision, gonzalez2024highly}.


This work aims to investigate a class of models where the potentials admit long-range kink solutions. Due to their strong interactions over a large distance, analyzing the dynamics of long-range kinks presents unique challenges and demands specialized methods for setting initial conditions. In particular, the influence of nearby kinks cannot be overlooked, making the simple additive ansatz inadequate, as shown in \cite{christov2019long}. The most accurate initial conditions for kink-antikink interactions to date were proposed in \cite{christov2019long, campos2021interaction}, requiring that the field and velocity field closely satisfy the static equation of motion and the zero mode equation, respectively. More recently, \cite{campos2024collision} proposed a simpler approach to initiate long-range kinks interactions using a half-BPS impurity. Although slightly less accurate, the method offers an efficient method with a negligible computational cost. In this work, we employ the same method to construct the moduli space for the collective coordinates approximation for long-range kinks' interactions, which is the first in the literature.

The outline of the paper is as follows. In Sect.~\ref{sec:col}, we revisit a class of models characterized by potentials with three minima, which admit kink solutions with one long-range tail. This section also explores the full dynamics of antikink-kink collisions in these models. In Sect.~\ref{sec:lin-stab}, we analyze the stability of long-range kinks by considering linear perturbations. Section~\ref{sec:delocalized}  investigates the pressure between the kink and antikink originating from the delocalized modes and the consequent impact on the dynamics. In Sec.~\ref{sec:CC}, we use the collective coordinate approximation to study the antikink-kink dynamics. We mainly focus on the $\phi^8$ model with one long-range and one short-range tail kink and also the $\phi^6$ model where the kink solution is short-range on both sides. We compare the results of the dynamics via collective coordinates with the ones obtained from full dynamics. The last section is devoted to summarizing the main results and concluding remarks. 

\section{Antikink-kink Collisions Revisited}
\label{sec:col}
\begin{figure}
    \centering
    \includegraphics[width=\textwidth]{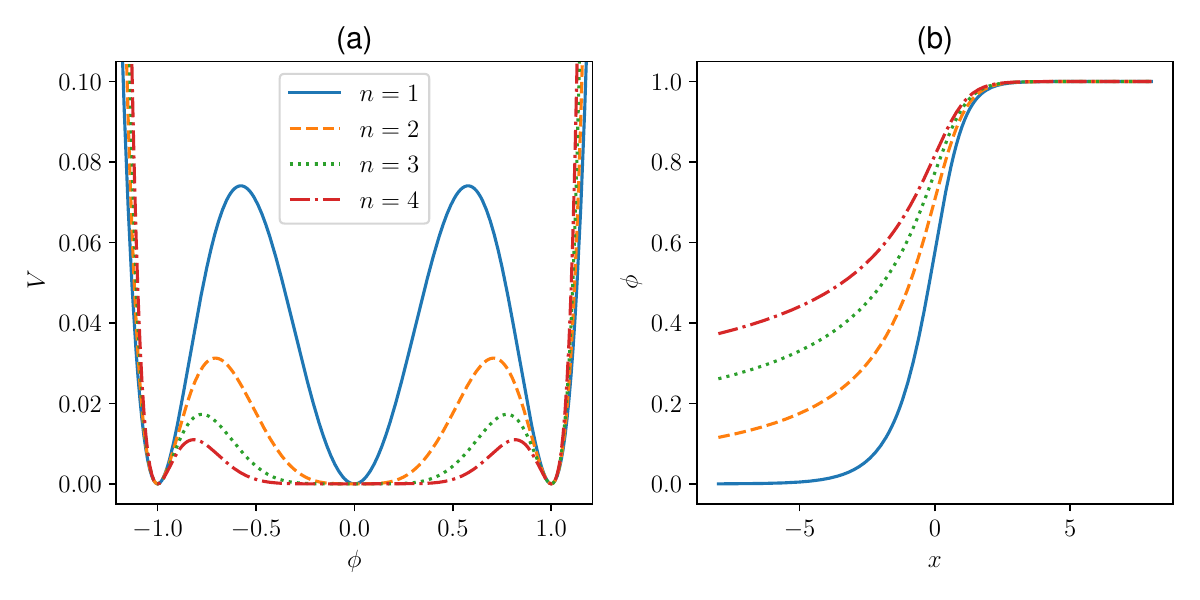}
    \caption{Potential and kink profiles for several values of $n$.}
    \label{fig:kink-n-pot}
\end{figure}

We consider the following scalar field theory in (1+1) dimensions
\begin{equation}
    \mathcal{L}=\frac{1}{2}\partial_\mu\phi\partial^\mu\phi-\frac{1}{2}\phi^{2n}(\phi^2-1)^2,
\end{equation}
with kink solutions interpolating between the three minima 0 and $\pm 1$ for the static field. The potential $U(\phi)$ and the kink $\phi_K(x)$ are exhibited in Fig.~\ref{fig:kink-n-pot}. The tail facing the minimum $0$ is long-range for $n\geq 2$ with asymptotic behavior $x^{-1/(n-1)}$. On the other hand, the tail facing the minima $\pm 1$ is short range and decays as $e^{-2x}$. The static equation of motion satisfies the BPS condition
\begin{equation}
  \phi_x=\pm W(\phi),
\end{equation}
where $W(\phi)=|\phi|^n(1-\phi^2)$. The solutions can be written in an implicit form 
\begin{equation}
    x-x_0=\int_{\phi_0}^{\phi_K}\frac{d\phi}{W(\phi)}\, .
\end{equation}
The integrals are
\begin{equation}
\phi_8^{-1}=\frac{1}{2}\ln\frac{1+\phi_K}{1-\phi_K}-\frac{1}{\phi_K},\quad\phi_{10}^{-1}=\frac{1}{2}\ln\frac{\phi_K^2}{1-\phi_K^2}-\frac{1}{2\phi_K^2},\quad\phi_{12}^{-1}=\frac{1}{2}\ln\frac{1+\phi_K}{1-\phi_K}-\frac{1}{\phi_K}-\frac{1}{\phi_K^3}.    \label{inveqs}
\end{equation}
The offsets are set at the position where the energy density admits maximum \cite{manton2019forces}. They can be found by demanding that $\phi''(x_0)=0$, which leads to the condition 
\begin{equation}
    \left.\frac{\partial U}{\partial\phi}\right|_{\phi=\phi(x_0)}=0\Rightarrow \phi(x_0)=\sqrt{\frac{n}{n+2}}
\end{equation}
Plugging it into \ref{inveqs}, one obtains the offsets
\begin{equation}
    x_0=\phi^{-1}_{4+2n}\left(\sqrt{\frac{n}{n+2}}\right).
\end{equation}

\begin{figure}
    \centering
    \includegraphics[width=0.9\textwidth]{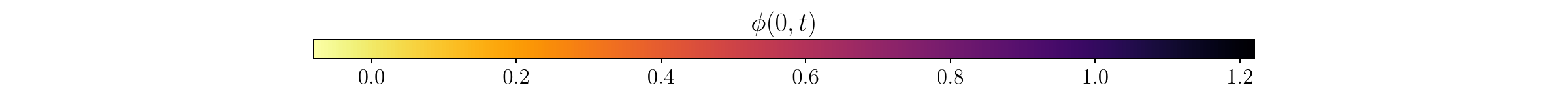}
    \includegraphics[width=0.9\textwidth]{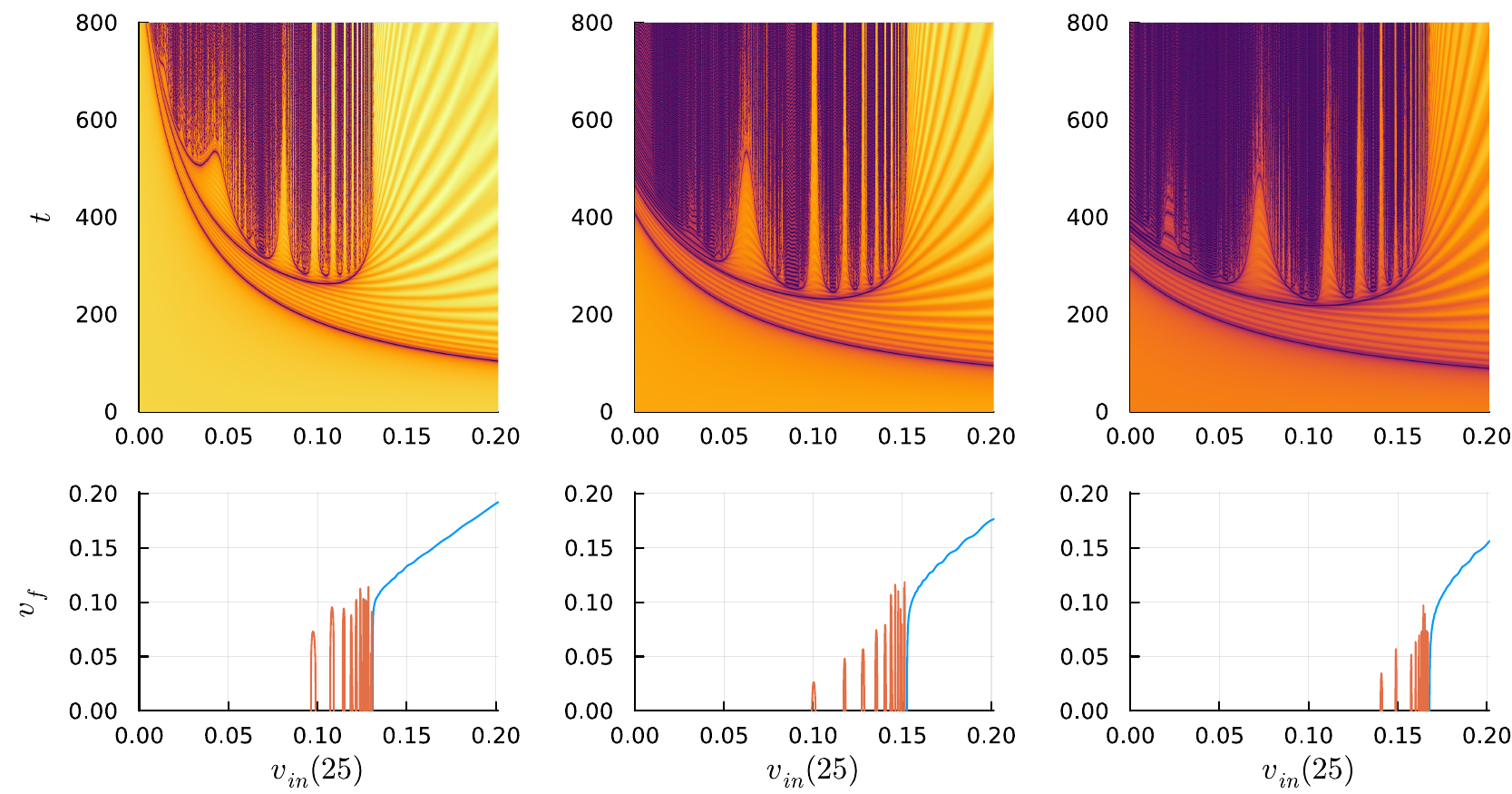}
    \caption{(Top row) Field at the center of mass as a function of time and initial velocity. (Bottom row) Final velocity of one- and two-bounce windows as a function of the initial velocity.}
    \label{fig:windows}
\end{figure}

There are two types of kink-antikink collisions in the present model. The tail that faces the opposing kink is either long-range or short-range. If it is short-range, the mass gap of the vacuum sector is zero. In the current setting, it implies that there is no resonance nor bion formation \cite{bazeia2023kink}. We are interested in the sector with a nonzero mass gap. More precisely, if we choose the kinks to obey $\phi\geq 0$, such a sector corresponds to an antikink-kink collision. As shown in Ref.~\cite{christov2021kink}, those models exhibit resonance windows with a quasi-fractal structure. We reproduce such scattering output in Fig.~\ref{fig:windows}. The present work aims to elucidate the mechanism behind such a resonance phenomenon.

The result in Fig.~\ref{fig:windows} was obtained using the specialized methods described in Ref.~\cite{campos2021interaction}. In short, one needs to construct the antikink-kink configuration first via the split-domain ansatz with separation $2x_0$, then minimize it according to the equations of motion for a traveling wave $u(x;x_0,v_{in})$ with velocity $v_{in}$, keeping the separation fixed. To construct the initial condition for $\dot\phi$, we start from the initial guess
\begin{equation}
    \dot\phi(x,0)=v_{in}\text{sgn}(x)\frac{du}{dx}.
\end{equation}
Then, we find the field configuration that obeys the Lorentz contracted zero mode equation as closely as possible, keeping the field fixed near the kinks center. 

\begin{figure}
    \centering
    \includegraphics[width=0.9\textwidth]{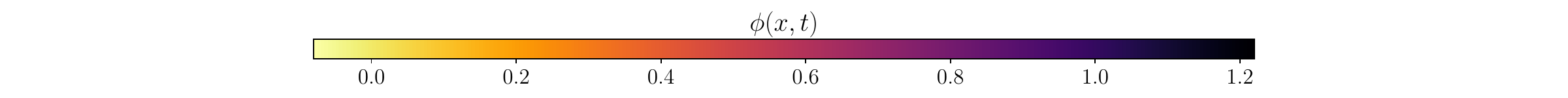}
    \includegraphics[width=0.96\textwidth]{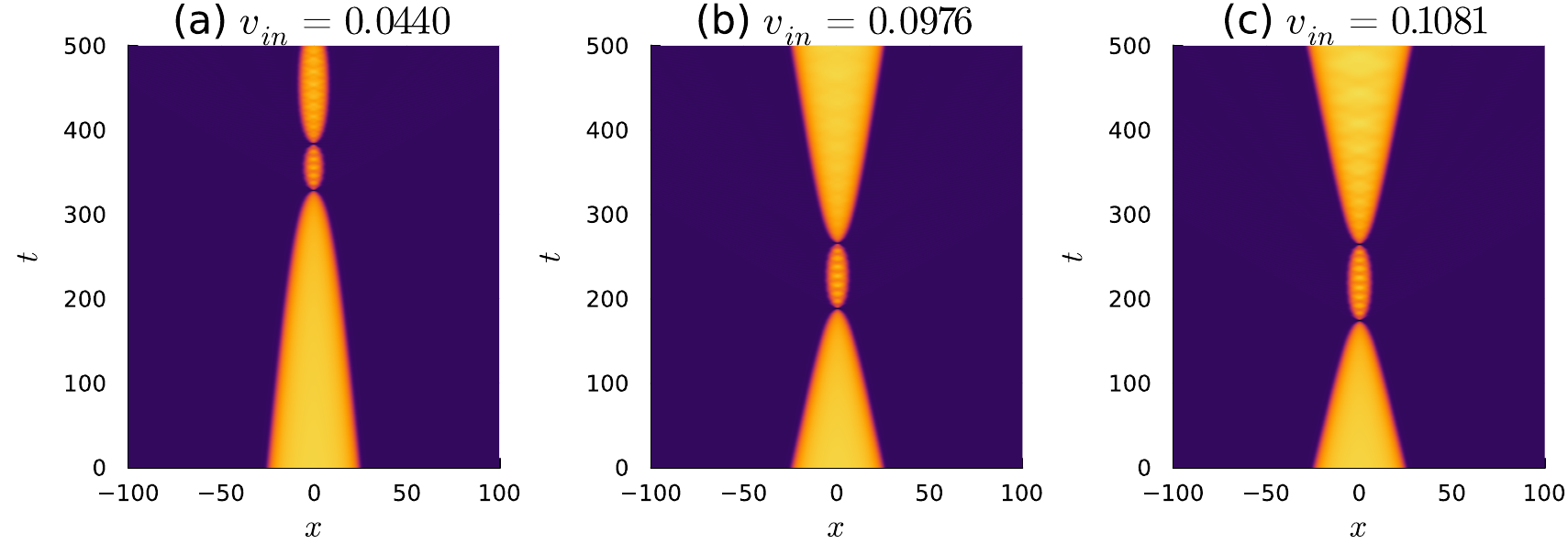}
    \caption{Spacetime evolution of $\phi$ in antikink-kink collisions of $\phi^8$ theory. The left figure corresponds to the first resonance window, which is a false one. The middle and right figures correspond to the second and third windows, which are true ones. The kinks are initially located at $x_0=25.0$.}
    \label{fig:cases}
\end{figure}

Interestingly, this model presents very large false resonance windows, which are intervals of initial velocities where the kinks acquire a large amount of energy at the second bounce but not enough to fully separate. Comparing false, Fig.~\ref{fig:cases}(a), and true windows, Figs.~\ref{fig:cases}(b) and (c), we see that the kink trajectory slowly decelerates and reverses its motion in false windows, while it reaches an asymptotic constant value in true resonance windows. 

It is important to mention that, as shown in Ref.~\cite{christov2021kink}, the scattering output for a given initial velocity depends on the choice of the initial position $x_0$ due to the power-law decay of the interaction force between the kinks. If $x_1$ and $x_2$ are sufficiently large, the initial velocities at these two positions can be mapped by the following relation
\begin{equation}
    \frac{1}{2}v_{in}^2(x_1)-\frac{\alpha_{4+2n}}{x_1^{(1+n)/(n-1)}}=\frac{1}{2}v_{in}^2(x_2)-\frac{\alpha_{4+2n}}{x_2^{(1+n)/(n-1)}},
\end{equation}
where $\alpha_{4+2n}$ is the proportionality constant of the potential energy between long-range antikink-kink pairs divided by the mass. From the results in Refs.~\cite{manton2019forces, christov2019kink}, we obtain $\alpha_8=3.693$. For the $\phi^8$ theory, we measure the critical velocity as $v_c(25)=0.131$. Thus, the critical velocity would be $v_c(16)=0.136$ for the initial position $x_0=16$. This information will be important below.

\section{Linear perturbation analysis}
\label{sec:lin-stab}

Let us consider perturbations around (momentarily) static antikink-kink configurations $u(x; x_0, v_{in}=0)$. The configurations are obtained as described in Ref.~\cite{christov2019long}. We start with the split domain ansatz and then find the configuration that best obeys the static equations of motion while fixing the interkink distance $2x_0$. 

\begin{figure}
    \centering
    \includegraphics[width=\textwidth]{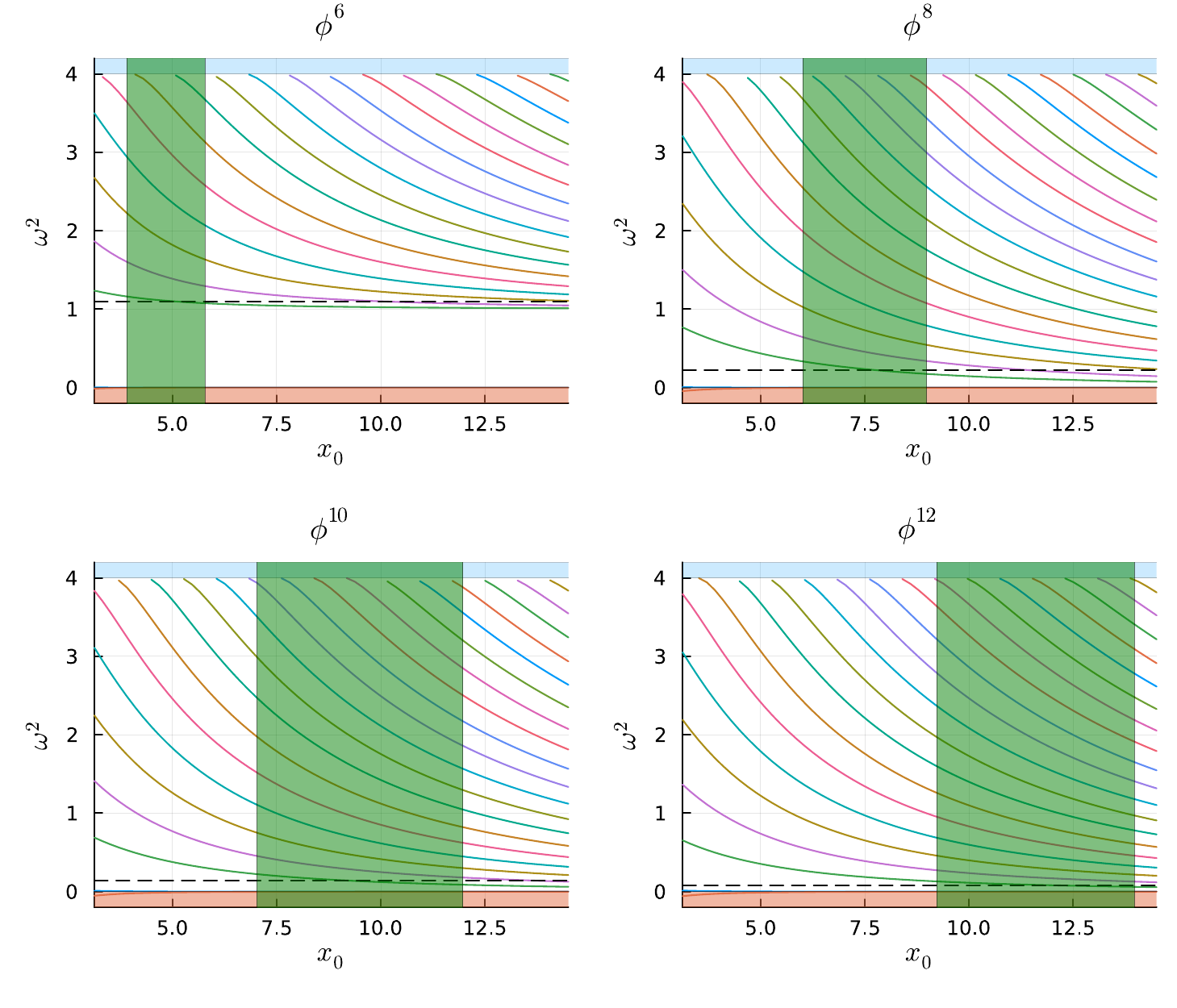}
    \caption{Spectrum of linear perturbations around antikink-kink configurations as a function of the half interkink distance $s_0$.}
    \label{fig:spec}
\end{figure}

The perturbations are obtained as usual by writing $\phi(x,t)=u(x;x_0, v_{in}=0)+e^{i\omega t}\eta(x)$. It yields the following Schr\"{o}dinger-like equation
\begin{equation}
\label{eq:lin-eta}
    \omega^2_k\eta_k(x)=\left[-\frac{d^2}{dx^2}+U^{\prime\prime}(u(x))\right]\eta_k(x),
\end{equation}
where the parameters in $u$ have been omitted and $k$ labels the eigenfunctions. The bound frequencies as a function of $x_0$ are exhibited in Fig.~\ref{fig:spec}. There are two near-zero modes, one stable $\eta_2$ and one unstable $\eta_1$. The instability is expected due to the antikink-kink force. The remaining ones are known as delocalized vibrational modes. A family of such modes exists, which becomes larger as $x_0$ increases. 

In Fig.~\ref{fig:spec}, vertical green stripes show the range of the largest interkink distances between consecutive bounces. The lowest end corresponds to the first true two-bounce resonance window, and the highest one corresponds to the highest-order two-bounce window encountered. We have searched for windows on the initial velocity interval $[0.0,0.2]$ with steps $\Delta v_{in}=0.0002$. For the $\phi^6$ theory, the highest end was considered to be the value reported in Ref.~\cite{dorey2011kink} for the maximal separation at the $109$th window. Notice that the range of separations is larger for the long-range range interaction ($n>1$) in comparison to the short-range case ($n=1$).

\begin{figure}
    \centering
    \includegraphics[width=\textwidth]{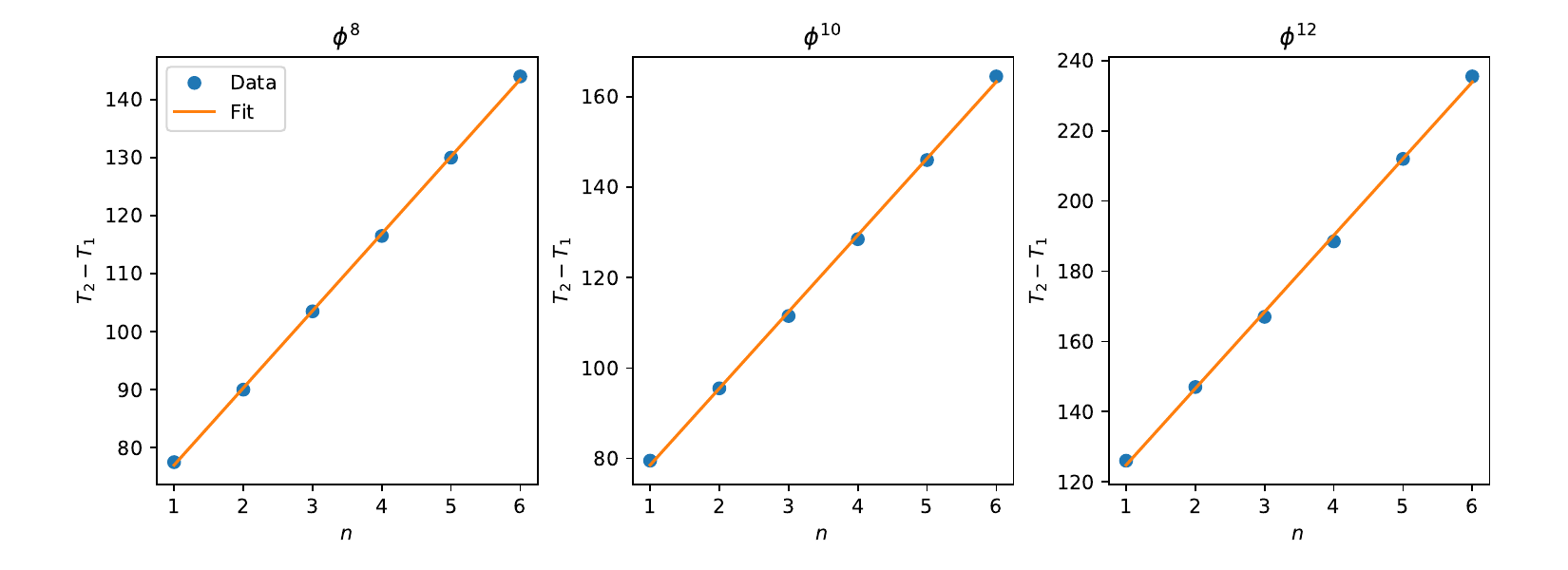}
    \caption{Time between bounces for the first six resonance windows. They are well fitted by a straight line.}
    \label{fig:freqs}
\end{figure}

The frequency responsible for the resonant energy exchange mechanism can be obtained from the simulated scattering output, as described in Ref.~\cite{campbell1983resonance}. First, we measure the time between the first and second bounce $T$ at the center of the $m$-th resonance window. Then, we fit the result according to the relation
\begin{equation}
\label{eq:Tm}
\omega T=2\pi m+\delta.
\end{equation} 
Our numerical calculations are shown in Fig.~\ref{fig:freqs}. They give the resonant frequencies $\omega_8\simeq 0.472$, $\omega_{10}\simeq 0.371$, and $\omega_{12}\simeq 0.288$, which are marked as horizontal lines in Fig.~\ref{fig:spec}. In all four cases, from $\phi^6$ up to $\phi^{12}$, the horizontal dashed line only crosses the lowest delocalized mode $\eta_3$, marked in green, in the range of maximal separations. Our construction, complementary to the one provided in Ref.~\cite{dorey2011kink}, indicates that $\eta_3$ is the mode responsible for the resonant energy exchange mechanism. We should mention that, if high-order windows were used here instead, the wide variation in the maximal separation would slightly decrease the measured value of the frequency.

\section{Delocalized mode pressure and wobbling kink collisions}
\label{sec:delocalized}

\begin{figure}
    \centering
    \includegraphics[width=0.65\textwidth]{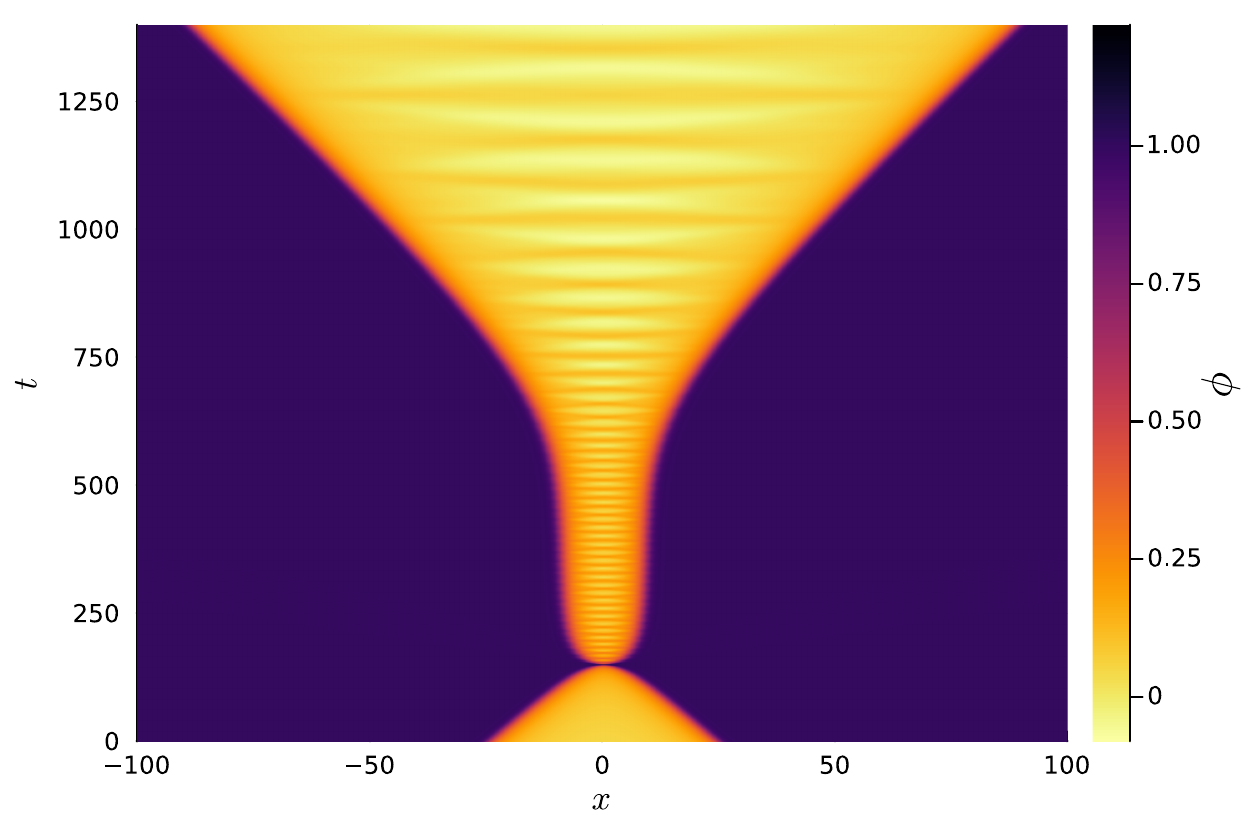}
    \caption{Evolution of the scalar field $\phi$ in spacetime for $v_{in}(25)=0.131$ for the $\phi^8$ model. It corresponds to the critical velocity.}
    \label{fig:critcol}
\end{figure}

A fascinating feature of the long-range kink collisions under investigation is the delocalized mode pressure. As shown in Fig.~\ref{fig:critcol}, evolving the field equations precisely at the critical velocity gives rise to two phenomena. First, as expected, the kinks collide and emerge with nearly vanishing velocity. However, the trapped modes in the inter-kink region subsequently exert pressure on the kinks, converting vibrational energy back to translational one.
As shown in Ref.~\cite{forgacs2008negative}, radiation exerts a positive pressure when its reflection coefficient is nonzero or when the radiation scatters from higher to lower mass region \cite{Romanczukiewicz:2008hi, Romanczukiewicz:2017hdu, Forgacs:2013oda}. When the kinks are far apart, the delocalized modes mostly resemble radiation.
 Therefore, they exert positive pressure on the kinks due to the mass difference between the vacua.

\begin{figure}
    \centering
    \includegraphics[width=0.65\textwidth]{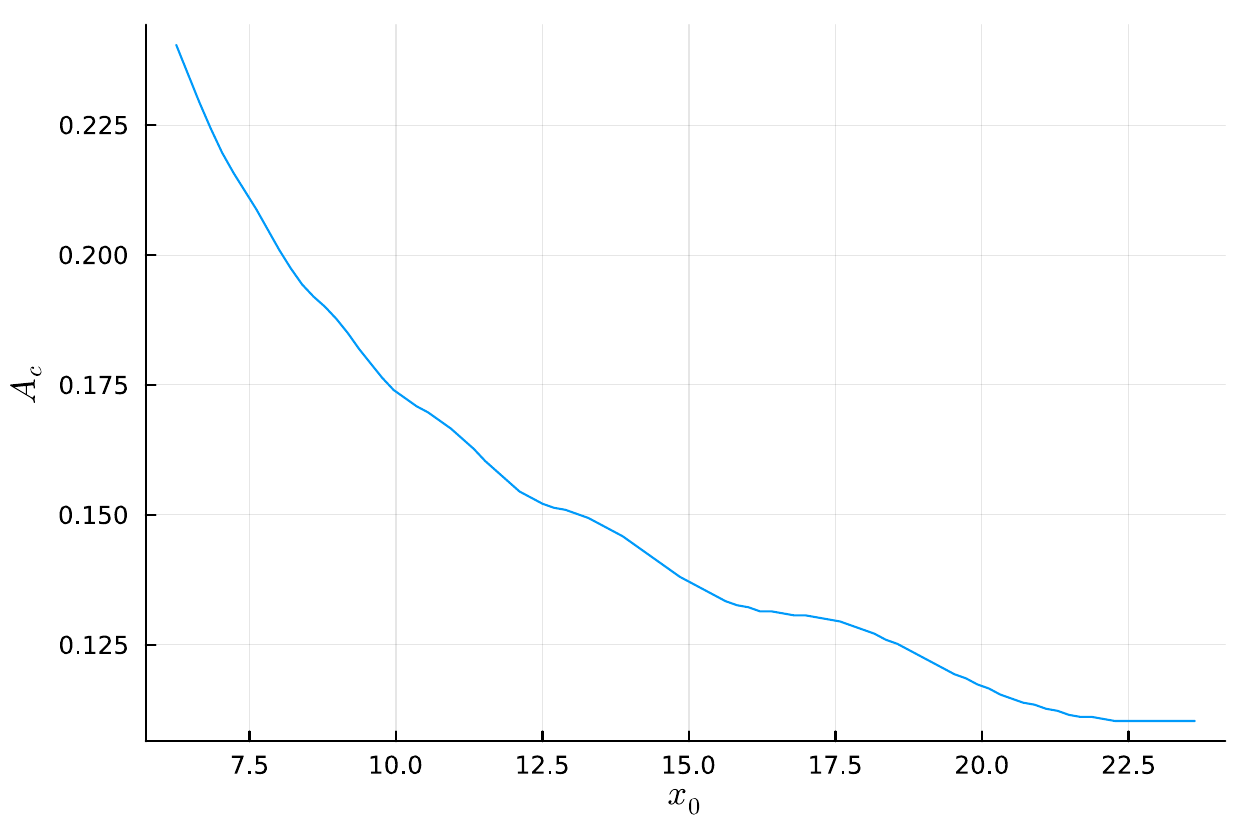}
    \caption{Critical amplitude $A_c$ of wobbling kink collisions as a function of the kink's starting position $x_0$ for the $\phi^8$ model.}
    \label{fig:critamp}
\end{figure}

As the interaction of wobbling kinks, where the vibration comes from a delocalized mode, has not been studied in the literature, we aim to study them here. We will see that the delocalized mode pressure, as described above, is a crucial aspect of such a physical process. Notice that one can construct a vibrating antikink-kink configuration that neither collapses nor separates by tuning the delocalized mode amplitude. This phenomenon occurs because the kinks' attractive force counterbalances the delocalized modes' positive pressure. One could call such a configuration a \textit {dynamical sphaleron}, as it is an unstable dynamical configuration that separates the two behaviors. The critical amplitude as a function of the kink's starting position $x_0$ is shown in Fig.~\ref{fig:critamp}. For lower $x_0$, the inter-kink force increases, requiring a larger amplitude to be compensated. It is essential to mention that the same construction can be employed in short-range models, such as $\phi^6$, thus having a general character.

\begin{figure}
    \centering
    \includegraphics[width=0.65\textwidth]{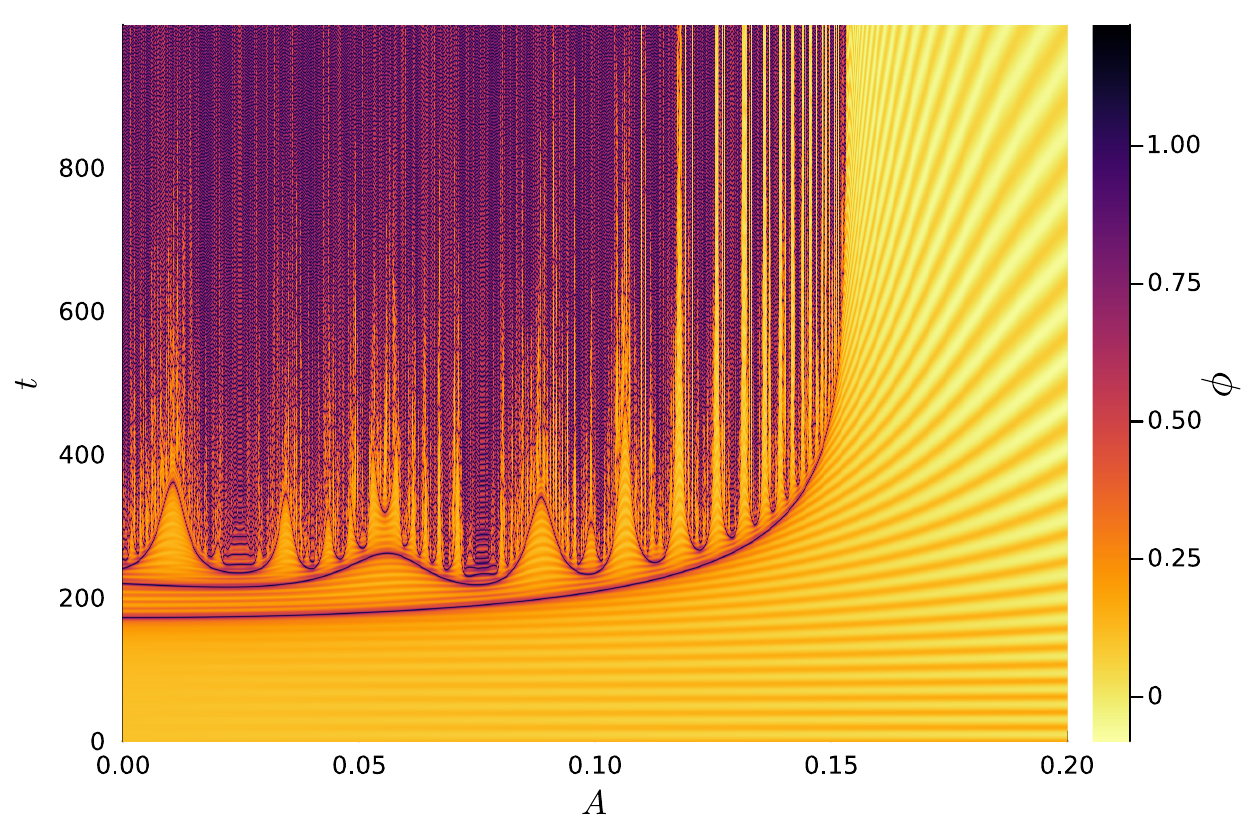}
    \caption{Field at the center of mass as a function of time and initial amplitude for the $\phi^8$ model.}
    \label{fig:phiAt}
\end{figure}

The numerical setup for Figs.~\ref{fig:critamp} and \ref{fig:phiAt} is the following. We start the simulations with $\dot\phi=0$ and 
\begin{equation}
    \phi(x,t)=u(x;x_0=12.5,v_{in}=0)+A\eta_3(x).
\end{equation}
Then, the fields evolve according to the equations of motion. The solution $\eta_3$ is obtained using the NDEigensystem function in Mathematica.\footnote{When computing $\eta_3$, we considered the kink-impurity with three parameters discussed in Ref.~\cite{campos2024collision} for the antikink-kink configuration.}

As shown in Ref.~\cite{adam2021sphalerons}, a sphaleron containing a vibrational mode can create resonant behavior. Similarly, here, near the critical amplitude, it is possible to obtain resonant behavior. In Fig.~\ref{fig:phiAt}, we show a nested set of resonance windows accumulating near the critical amplitude. The construction above is precisely the analog of wobbling kinks collisions for delocalized vibrational modes \cite{alonso2021scattering}. Remarkably, our results show that wobbling kink collisions via delocalized modes exhibit similar behavior to the ones of $\phi^4$ wobbling kinks.

\section{Collective coordinates Effective Model}
\label{sec:CC}

Our task here is to find a suitable moduli space for the antikink-kink collisions. The idea behind the collective coordinate method is to reduce the field to a set of configurations parameterized by a finite set of time-dependent variables $A^i(t)$, called moduli, in the form
\begin{equation}
    \phi(x,t)=\Phi(x; \{ A^i \} ).
\end{equation}
Then, substituting it in the original Lagrangian density and integrating over $x$, we obtain the following effective Lagrangian
\begin{equation}
    L_{\text{eff}}=g_{ij}(\{A^i\})\dot{A}^i\dot{A}^j-V(\{ A^i\}),
\end{equation}
where the metric $g_{ij}$ and effective potential $V$ are given by
\begin{align}
    g_{ij}(\{A^i\})&=\frac{1}{2}\int \frac{\partial\Phi}{\partial A^i}\frac{\partial\Phi}{\partial A^j}dx\\
    V(\{A^i\})&=\int \frac{1}{2}\left(\frac{\partial\Phi}{\partial x}\right)^2-U(\Phi(x,\{A^i\}))dx.
\end{align}
Equipped with the effective Lagrangian, one can obtain the equations of motion.

To investigate whether the resonance exchange mechanism applies here, the moduli space should contain at least two parts. Here, we are considering the antikink-kink configuration $\phi_{AKK}$ and one Derrick mode $\tilde\eta_D$ \cite{adam2022relativistic}. The Derrick mode will simulate the Lorentz factor and also partially the delocalized bound modes. We will discuss this point in more detail later. Therefore, we will see below that an appropriate moduli space is given by
\begin{equation}
\label{eq:MS}
\phi_{CC}(x;\{A^i(t)\})=\phi_{AKK}(x;A^1(t))+A^2(t)\tilde\eta_D\left(x;\left[A^1(t)\right]^2\right),
\end{equation}
where $A^1$ is the modulus representing the kink position, while $A^2$ represents the amplitude of the resonant mode $\tilde\eta_D$. We chose $A^1(t)$ with the squared as the argument in $\tilde\eta_D$ for the reason that will become clear shortly.

To evolve the equations of motion, suitable initial conditions should be given. We can fix the kink position and velocity by appropriately choosing $A^1$ and $\dot{A}^1$. Moreover, as shown in Ref.~\cite{manton2021collective}, it is important to consider the Lorentz contraction of a moving kink. Otherwise, we obtain a wobbling kink instead. Mathematically, it implies that $\ddot{A}^2=\dot{A}^2=0$, which can be solved numerically for $A^2(0)$. Thus, it only remains to find $\phi_{AKK}$ and $\tilde\eta_D$.

\subsection{Impurity ansatz for the antikink-kink configurations}

The antikink-kink configuration will be constructed from the impurity ansatz. The idea behind the ansatz consists of building multiple kink configurations from the BPS equation of the original theory with the addition of a half-BPS preserving impurity \cite{manton2019iterated, adam2020kink}. In Ref.~\cite{campos2024collision}, it has been proposed as the initial conditions of kink collisions and proven efficient, especially in the long-range regime.

\begin{figure}
    \centering
    \includegraphics[width=\textwidth]{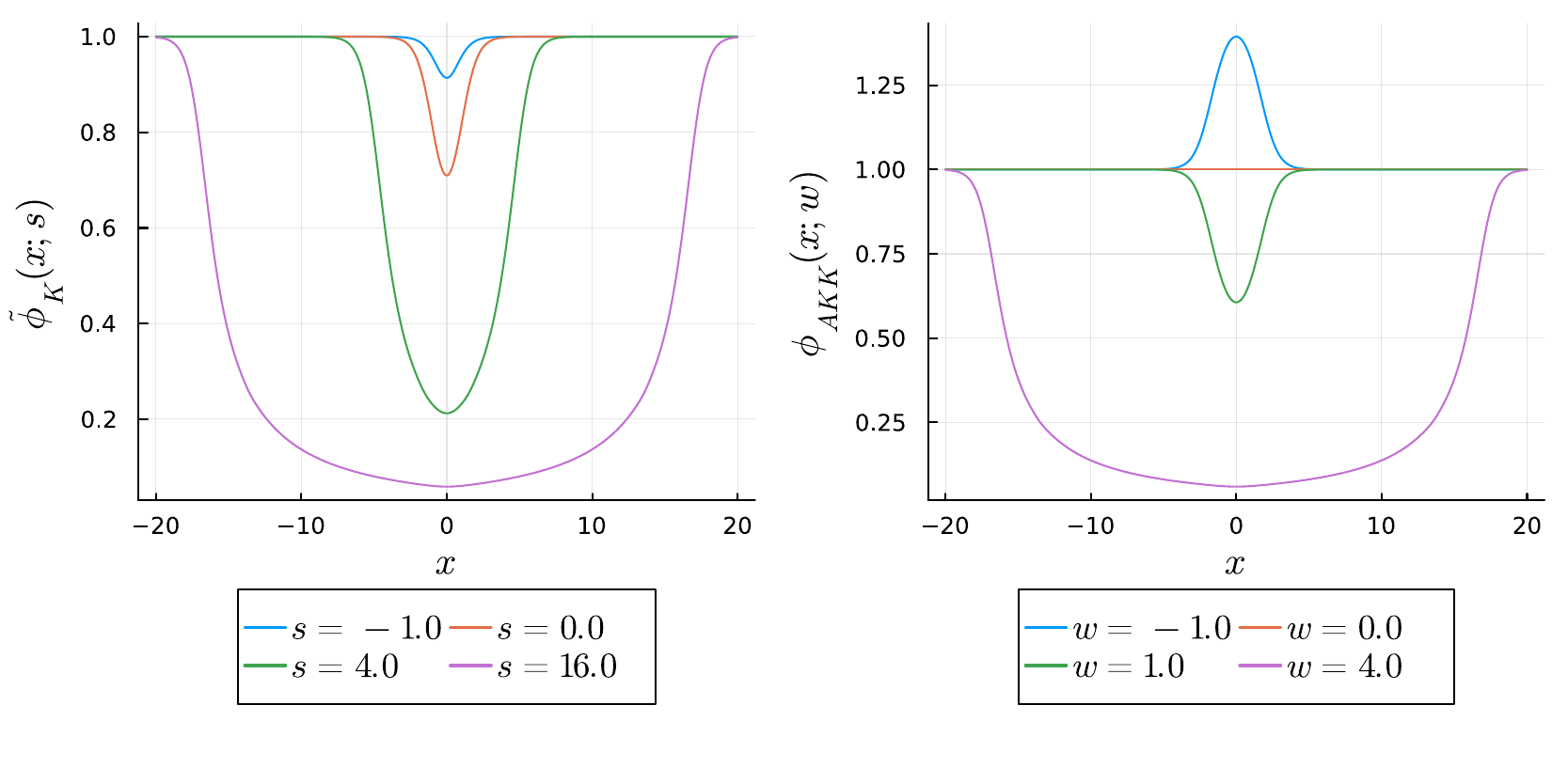}
    \caption{Antikink-kink approximate configurations for the $\phi^8$ model. We exhibit two cases: (left) $\tilde\phi_K(x,s)$ and (right) $\phi_{AKK}(x;w)$.}
    \label{fig:moduli}
\end{figure}

Let us consider the following half-BPS preserving field theory
\begin{equation}
    \mathcal{L}=\frac{1}{2}\tilde\phi_t^2-\frac{1}{2}\left(\tilde\phi_x-\sigma(x)W(\tilde\phi)\right)^2.
\end{equation}
Then, the BPS equation becomes
\begin{equation}
    \tilde\phi_x=\sigma(x)W(\tilde\phi)
\end{equation}
We consider the impurity $\sigma(x)=\tanh(x)$, leading to profiles with an antikink-kink character. The impurity contribution is analytically solvable \cite{adam2019solvable}. We define a new variable
\begin{equation}
\xi(x;s)=\log(\cosh x)-s.
\end{equation} 
Then, in terms of a single kink profile $\phi_K(x)$, the solution becomes 
\begin{equation}
    \tilde\phi_{K}(x;s)=\phi_K(\xi(x;s)),
\end{equation}
This solution exhibits an antikink-kink structure, as illustrated in Fig.~\ref{fig:moduli}(a). Notably, for large values of $s$, the antikink and kink are centered at $-s$ and $s$, respectively. In other words, as $s$ increases, it asymptotically approaches $x_0$.

\subsection{The generalized Derrick mode}

\begin{figure}
    \centering
    \includegraphics[width=\textwidth]{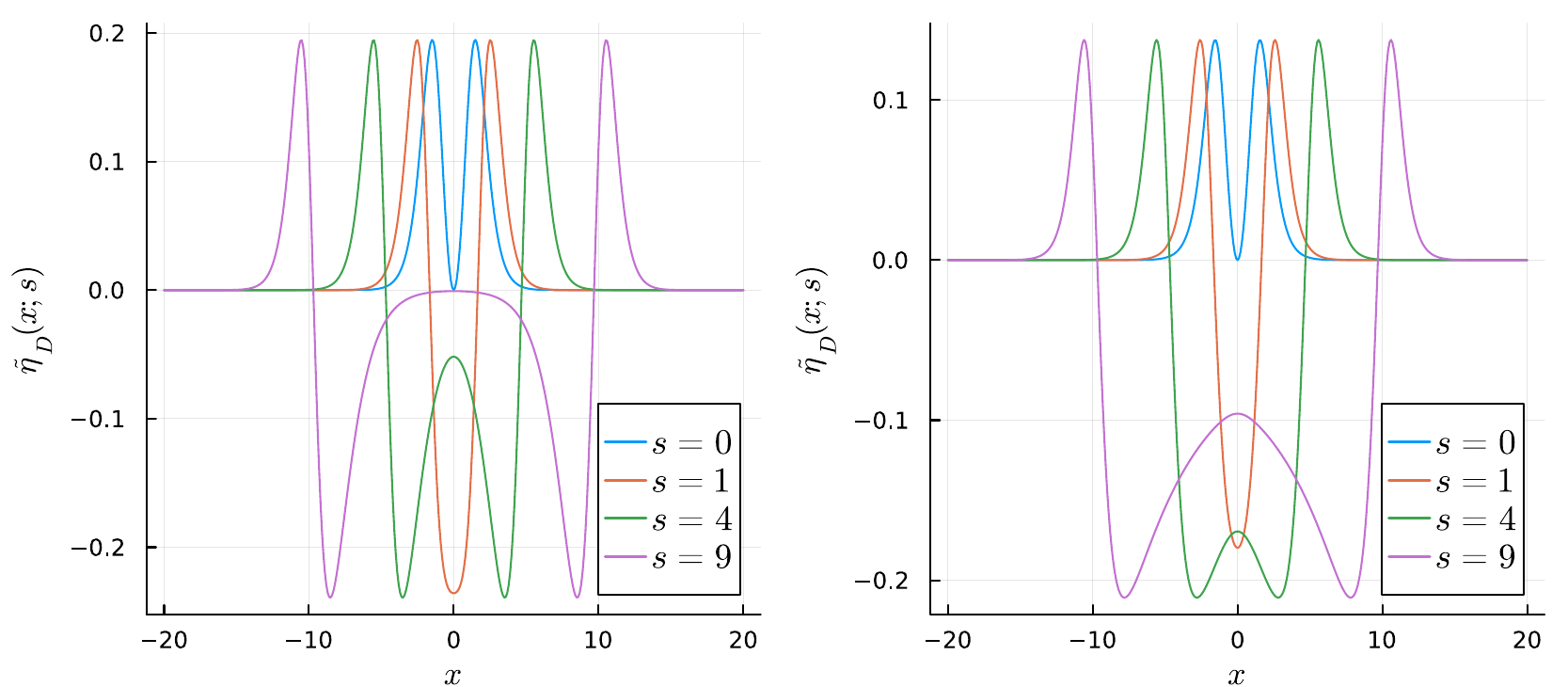}
    \caption{Profile of the generalized Derrick mode for several values of $s$. The models are $\phi^6$ (left) and $\phi^8$ (right).}
    \label{fig:gen-derrick}
\end{figure}

The next step in the moduli space construction is to include the Derrick mode $\eta_D(x)=x\phi_K^\prime(x)$. It is necessary for dynamical phenomena involving kinks because it is a perturbative way of incorporating the Lorentz contraction \cite{adam2022relativistic}. As shown in Ref.~\cite{campos2024collision}, the impurity ansatz consists essentially in setting $x\to \xi$ in both $\phi_K(x)$ and the zero mode, $\phi_K^\prime(x)$. A natural generalization, therefore, is to consider the following form for the Derrick mode
\begin{equation}
    \tilde\eta_D(x;s)=\eta_D(\xi(x;s))=\xi(x;s)\phi_K^\prime(\xi(x;s)).
\end{equation}
Its profile is shown in Fig.~\ref{fig:gen-derrick} for both $\phi^6$ and $\phi^8$ model.

It is important to mention that, for full theory simulations involving long-range kinks, it is necessary to include some parameters in the impurity profile and find their optimal values \cite{campos2024collision}. However, we will ignore this step here for simplicity. Fortunately, the effective model reproduces the resonant structure fairly well without including any parameter, highlighting its robustness. Another important observation is that correctly centering the kink at the inflection point is a crucial step in defining the Derrick mode and obtaining the properties described below.

Consider the linearized problem eq.~\eqref{eq:lin-eta} with the approximation $\tilde\phi_K(x;s)$ instead of $u$. This last step is important to align the position of the kinks and $\tilde\eta_D$. Let us define the overlap between normalized linearized solutions by
\begin{equation}
    \langle\tilde\eta_D,\eta_n\rangle=\int\tilde\eta_D(x;s)\eta_n(x;s)dx.
\end{equation}
Similarly, the frequency is computed as
\begin{equation}
    \omega_D^2(s)=\left\langle\tilde\eta_D\left|-\frac{d^2}{dx^2}+U^{\prime\prime}(\tilde\phi_K)\right|\tilde\eta_D\right\rangle,
\end{equation}
where the arguments of $\tilde\phi_K$ have been omitted. Interestingly, the generalized Derrick mode has two interesting properties: it has a large overlap with the lowest delocalized mode $\eta_3$ for small $s$ and a frequency in the interval $m_1^2<\omega^2<4$, where the lower and upper limits are the squared masses in inner and outer vacua of the potential, respectively. 

\begin{figure}
    \centering
    \includegraphics[width=\textwidth]{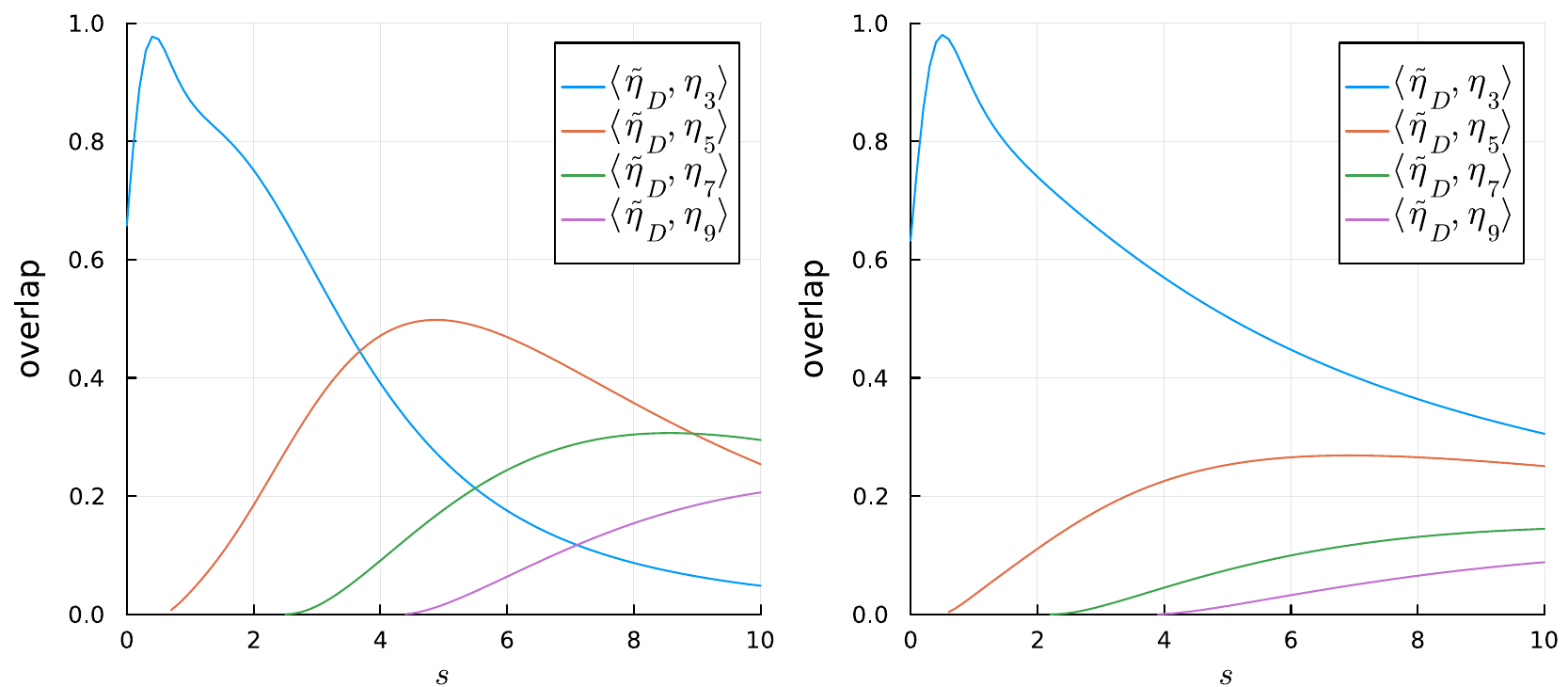}
    \caption{Overlap between $\tilde\eta_D$ and $\eta_n$ as a function of $s$ for several values of $n$.}
    \label{fig:overlap}
\end{figure}

\begin{figure}
    \centering
    \includegraphics[width=\textwidth]{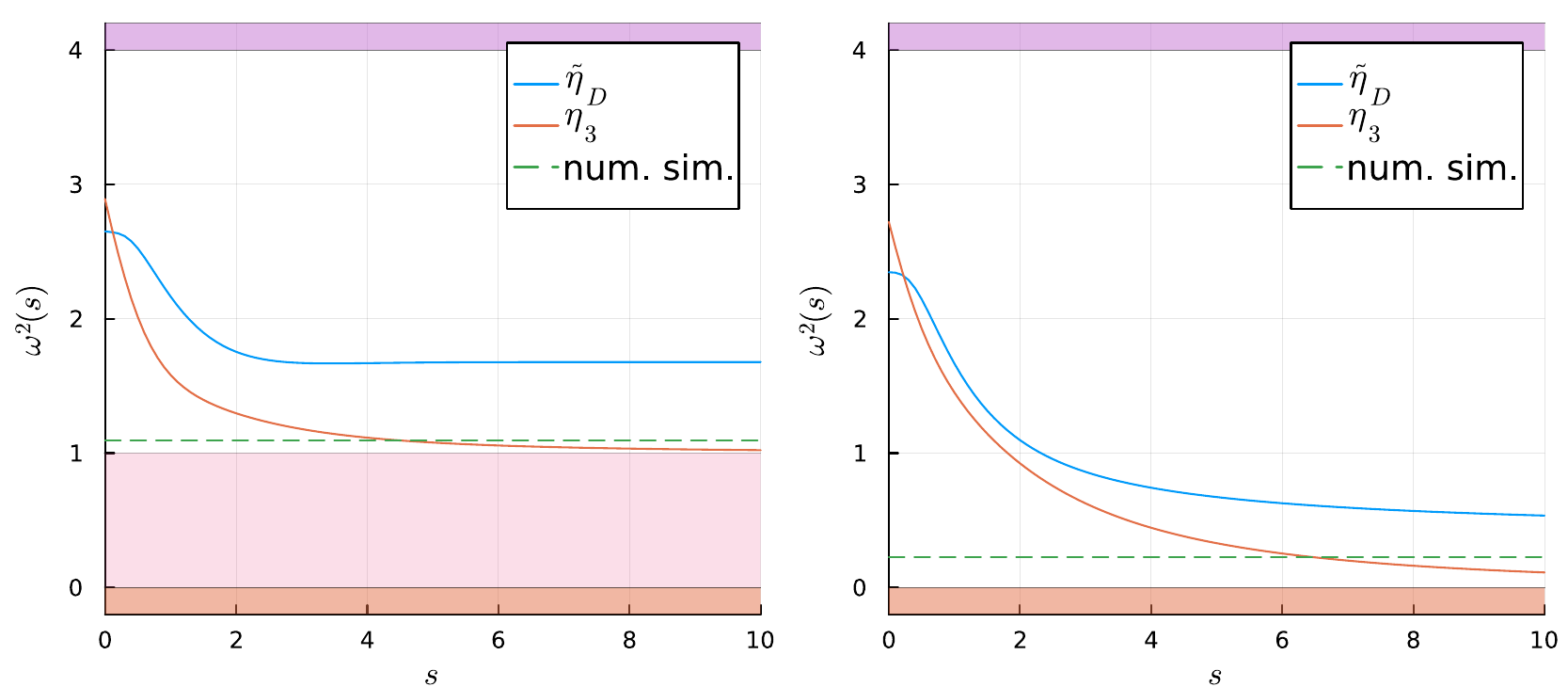}
    \caption{Frequencies $\omega_D^2$ and $\omega_3^2$ as a function of $s$. The numerical resonant frequency obtained from full theory scattering simulations is shown as a dashed line. Colored strips denote regions of continuum modes (purple), unstable modes (red), and localized modes (pink). We consider $\phi^6$ theory (Left) and $\phi^8$ theory (right).}
    \label{fig:freq-derrick}
\end{figure}

The overlap and the frequency are shown in Figs.~\ref{fig:overlap} and \ref{fig:freq-derrick}, respectively. Notice that the generalized Derrick mode appears as a mixture of several delocalized modes for large $s$. However, as $s$ decreases and the other modes transition into scattering states, the overlap exceeds 0.95. For the $\phi^8$, the generalized Derrick mode consists mainly of $\eta_3$, even when the kinks are further apart. Therefore, the generalized Derrick mode is a great candidate for the vibrational mode responsible for the resonant behavior, especially for the $\phi^8$ theory.

In Fig.~\ref{fig:freq-derrick}, we observe a shortcoming in the effective model. Namely, the generalized Derrick mode's frequency does not cross the frequency measured with full theory simulations. Therefore, the effective model cannot reproduce the correct resonant frequency, especially in the $\phi^6$ theory. Nevertheless, the two frequencies are not very far in the long-range case, and consequently, we have obtained a fairly reasonable set of resonance windows in Fig.~\ref{fig:scan8}. This evidences that effective models containing translational and vibrational modes are usually robust in reproducing resonant behavior.\footnote{See Ref.~\cite{kevrekidis2019four} for a historical example.}

\subsection{Removing the singularity}

The effective model, as described so far, contains a singularity and, thus, diverges when evolved numerically. The reason is that the solution $\tilde\eta_D$ is singular at the trivial vacuum solution, obtained at the limit $s\to-\infty$. This property can be verified by computing the moduli space's curvature at that point. Moreover, as $s$ decreases, the metric becomes vanishingly small, and the system reaches the singularity in a finite time.\footnote{A similar situation occurs in the kink-impurity system of Ref.~\cite{adam2023moduli}.} To remedy this problem, we need to modify the moduli space, avoiding the singularity. Hence, we change variables to $s=w^2$, leading to the functions $\tilde\eta_D(x;w^2)$ and $\tilde\phi_K(x;w^2)$ for the antikink-kink configuration and the vibrational mode, respectively. This choice reduces the moduli space, removing all configurations with $s<0$. In particular, it only describes configurations with $\phi<1$.  
Now, we can extend the moduli space by appropriately introducing a $\tanh(\beta w)$ factor. This way, we can access the region $\phi>1$, necessary to describe configurations near the bounce. We write the final antikink-kink configurations as
\begin{equation}
    \phi_{AKK}(x;w)=1-\tanh(\beta w)(\tilde\phi_K(x;w^2)-1).
\end{equation}
The profiles are shown in Fig.~\ref{fig:moduli}(b). They correctly mimic bouncing behavior for negative $w$ and are not singular at $w=0$, the vacuum solution. Finally, we set $w=A^1$ and substitute it in the expression in eq.~\eqref{eq:MS} to study the antikink-kink dynamics.

\subsection{Effective model for $\phi^6$ theory}

As a test of our moduli space, let us apply our construction to the $\phi^6$ model. First, we fix the parameter $\beta=10$. The particular value of $\beta$ is irrelevant, but it must be large. Otherwise, the field near $x=0$ moves away too fast from the vacuum at $\phi=0$, creating a repulsive force between the kinks.

\begin{figure}
    \centering

    \includegraphics[width=0.8\textwidth]{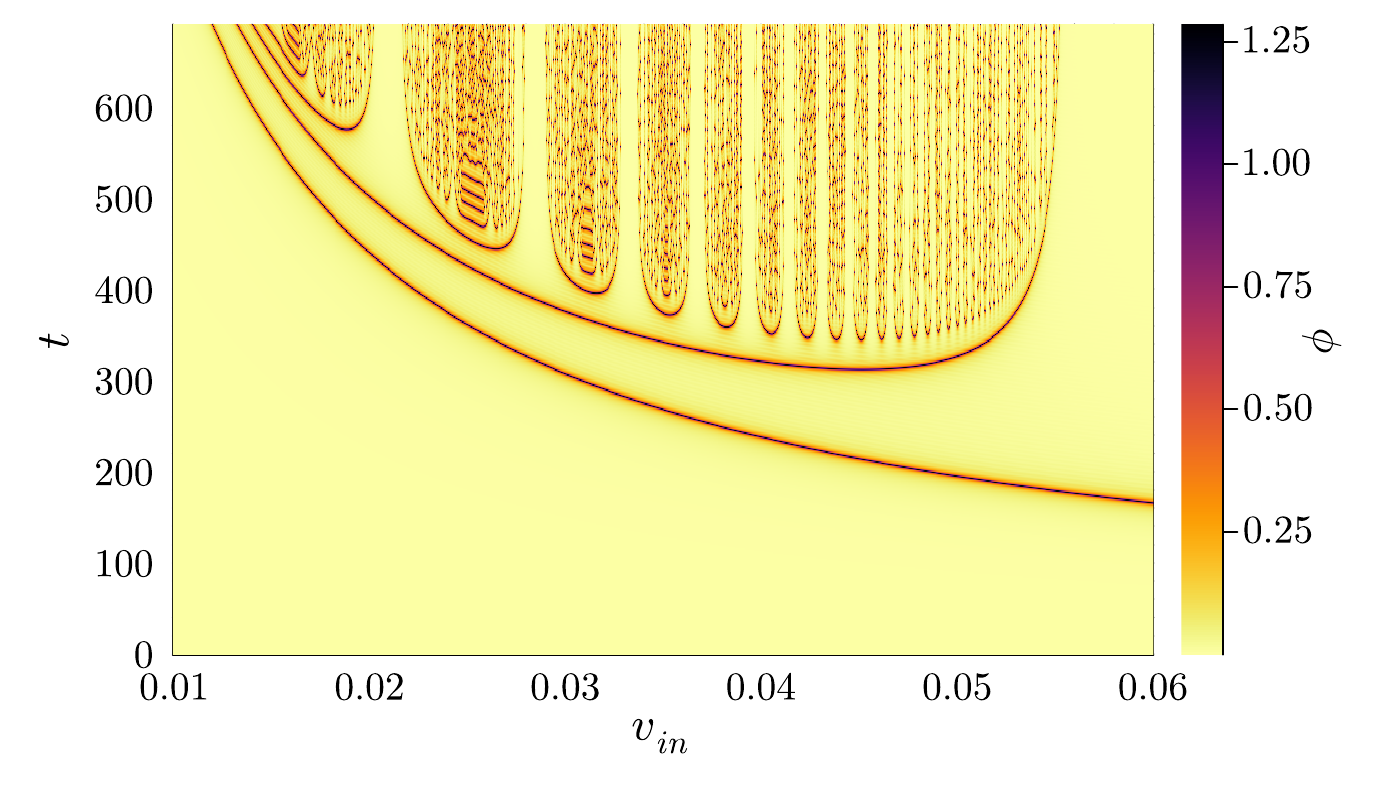}
    \caption{Field at the collision center as a function of $t$ and $v_{in}$. It is obtained from the $\phi^6$ theory effective model.}
    \label{fig:scan6}
\end{figure}

The effective model scattering output is illustrated in Fig.~\ref{fig:scan6}. Interestingly, it shows a sequence of two-bounce windows and a quasi-fractal structure of higher-bounce windows, but it does not reproduce the false resonance windows appearing in the full theory. The critical velocity is $v_c=0.0551$, which should be compared with the full theory $v_c=0.0457$ \cite{dorey2011kink}. The associated error is $21\%$, meaning that the effective model does not perform particularly well for the $\phi^6$ theory. 

We also computed the effective model's resonant frequency according to eq.~\eqref{eq:Tm}. We obtained $\omega_6^{eff}=1.812$, which should be compared with the theoretical frequency $\omega_6=1.045$ \cite{dorey2011kink}. The relative error is $e_6^{rel}=73\%$ and the absolute error is $e_6^{abs}=0.767$. The error is quite large, as expected from previous considerations. Although our effective model is less accurate than the best model available in the literature so far \cite{adam2022multikink}, it has the advantage that the qualitative behavior is reproduced in a minimal set containing only one translational degree of freedom and a Derrick mode.

\subsection{Effective model for $\phi^8$ theory}

Finally, we consider the $\phi^8$ model and test our moduli space for the kinks with long-range interactions. It is important to have an analytical expression for the kink profile for numerical reasons, but unfortunately, it does not exist. So, we proceed by constructing a fitting function in the form
\begin{equation}
    \phi_{\text{fit}}(x;\vec{a},\vec{b},\vec{c})=\epsilon(x;\vec{a})\phi_s(x;\vec{b})+(1-\epsilon(x;\vec{a}))\phi_l(x;\vec{c}),
\end{equation}
where $\epsilon(x;\vec{c})$ is an interpolation function, $\phi_s(x;\vec{c})$ is a short-range kink, and $\phi_l(x;\vec{c})$ is a long-range one. They all interpolate between 0 and 1 and contain a finite set of free parameters indicated by semicolons. Our choices are
\begin{align}
    \epsilon(x;\vec{c})&=\phi_s(x;\vec{c})=\frac{1+\tanh(c_1(x-c_2))}{2}\\
    \phi_l(x;\vec{c})&=\frac{1}{2}+\frac{\arctan(c_1(x-c_2))}{\pi}
\end{align}
After finding the optimal parameters, we obtain a maximum difference between $\phi_K$ and $\phi_{\text{fit}}$ of order $10^{-3}$. We also need a fit for $\phi_K^\prime(x)$. Similarly, the trial functions are constructed from the same choice of functions as
\begin{equation}
\eta_{\text{fit}}(x;\vec{a},\vec{b},\vec{c})=\epsilon(x;a_1,a_2)b_1\phi_s^\prime(x;b_2,b_3)+(1-\epsilon(x;a_1,a_2))c_1\phi_l^\prime(x;c_2,c_3).    
\end{equation}
After fitting, the maximum difference to $\phi_K^\prime(x)$ is or order $10^{-4}$.

\begin{figure}
    \centering
    \includegraphics[width=0.8\textwidth]{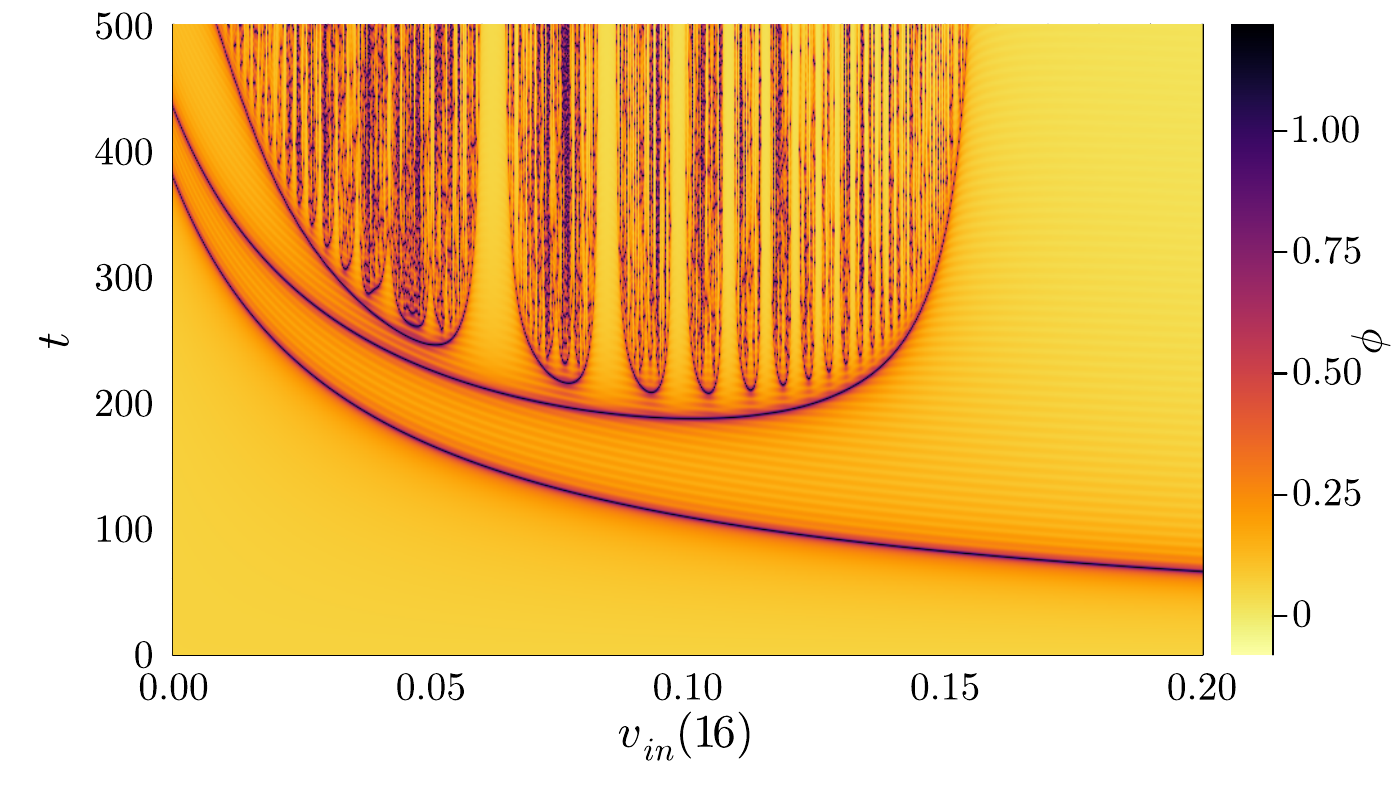}
    \caption{Field at the collision center as a function of $t$ and $v_{in}$. It is obtained from the $\phi^8$ theory effective model. Initial and final velocity are measured at $s_0=16$.}
    \label{fig:scan8}
\end{figure}

Once the fit has been obtained, the effective model can be integrated as described above. Due to the long-range character, the model is less restrictive concerning $\beta$, which is fixed to one. The result is shown in Fig.~\ref{fig:scan8} starting the kink at a separation $s_0=16$. Again, a nested set of resonance windows around a critical velocity is obtained. The critical velocity is $v_c=0.155$, corresponding to an error of 14\% compared to the value obtained in sec.~\eqref{sec:col}. As expected, the $\phi^8$ effective model performs better than $\phi^6$ one and has a more acceptable error. One of its shortcomings is that contrary to the full theory result in Fig.~\ref{fig:windows}, the lowest windows are not false in the effective model.

We have measured once more the effective model's resonant frequency according to eq.~\eqref{eq:Tm}. We obtained $\omega_8^{eff}=0.849$, which should be compared with the result from full theory simulations $\omega_8=0.472$. The relative error is $e_8^{rel}=80\%$ and the absolute error is $e_8^{abs}=0.377$. As expected from previous considerations, the error in the effective model resonant frequency is again quite large. Inspecting the relative error, it may seem that the effective model is less accurate for the $\phi^8$ model than the $\phi^6$ one. However, notice that the absolute error has decreased considerably. The increase in the relative error is due to the smallness of the theoretical value, which may lead to large relative errors from small absolute errors.  

To summarize, we have provided a setup with only two moduli that capture qualitatively the resonant behavior of the $\phi^8$ model. Hence, we have obtained further evidence that delocalized modes can mediate the resonance energy exchange mechanism. We have also shown how the delocalized mode may evolve dynamically as the kinks approach each other. Moreover, we provided a possible pathway in which the lowest delocalized mode $\eta_3$ is excited from the evolution of the Derrick mode.

\section{Conclusion}

In this paper, we investigated a class of models characterized by an integer $n$, in which the potentials have three minima. The kink solutions are short-range for $n=1$, corresponding to the $\phi^6$ model, i.e., exponentially tending to the vacuum values. For larger $n$, the middle minimum of the potential becomes massless. In this case, the kink solutions become long-range, power-law, in the tail facing the massless minimum. Besides the full dynamics, we studied the antikink-kink interactions using the effective model of collective coordinates. For the collective coordinates, we built upon two moduli, the kink position and the amplitude of a Derrick mode. For the initial antikink-kink configuration, we chose the kink-impurity ansatz with the desired character proposed in \cite{campos2024collision} instead of an additive ansatz, which is not a good initial guess for the kinks with long-range interactions as known in the literature. Taking this ansatz, it became essential to generalize the Derrick mode with the kink centered at the inflection point. We mainly focused on $n=2$, $\phi^8$ model, and compared the results with the $\phi^6$ one. We discussed that the Derrick mode is singular in the moduli space we constructed. Therefore, we had to circumvent the problem by cutting part of the moduli space and then extending it by a tanh function. This way, we could remove the singularity and access field configurations above and below the vacuum $\phi=1$.
Additionally, we projected the Derrick mode onto the bound delocalized modes and observed that, at large antikink-kink separations, the Derrick mode appears as a superposition of these delocalized modes.
As the separation distance decreases, the overlap with all modes diminishes, except for the lowest frequency mode, where overlap can exceed 0.95. We also analyzed the frequency. The Derrick mode frequency differs considerably from the lowest delocalized mode for large antikink-kink separation distance. The difference decreases by reducing the distance, as expected. However, the main issue is that the Derrick mode does not cross the frequency obtained from full dynamics for any antikink-kink separation distance, as shown in Fig.~\ref{fig:freq-derrick}. 
This discrepancy is particularly pronounced for the $\phi^6$ model, where the frequency difference is notably more significant.

The effective model dynamics for both the $\phi^6$ and $\phi^8$ theories exhibited strong qualitative agreement with the full dynamics, successfully capturing essential features such as resonance windows and critical velocities, aside from false windows. However, notable quantitative discrepancies were found in the resonant frequencies, as expected from the mismatch between the resonant and Derrick mode frequencies. We showed that one could also investigate the dynamics at the critical velocity. In this case, the kinks collide and re-emerge with near-zero final velocity. The trapped delocalized modes, which are similar to radiation, especially when the kinks are far apart, can create positive pressure in the region between the kinks. The trapped vibrational energy can be transformed back into the translational energy, boosting the velocity. This phenomenon is illustrated in Fig.~\ref{fig:critcol}.

The study of wobbling kinks' interactions, exciting a delocalized vibrational frequency, is lacking in the literature. To address this gap and explore the pressure exerted by delocalized modes on kinks in greater detail, we examined an antikink-kink configuration initially at rest, with the lowest delocalized mode excited. By adjusting the amplitude of the vibration, we constructed a wobbling antikink-kink configuration that neither decays nor reflects. In this scenario, the repulsive force from the delocalized modes counteracts the attractive static force between the kink and antikink. However, this critical amplitude depends on the initial separation of the kinks, and we referred to this configuration as a dynamical sphaleron. We showed the time evolution of the wobbling antikink-kink collision as a function of the delocalized bound-mode amplitude in Fig.~\ref{fig:phiAt}. The outcome closely resembles the one expressed in terms of the initial velocity. The vertical asymptote of the smooth curve that marks the first bounce indicates the critical amplitude. It separates the region where the kinks reflect from where resonance occurs.

While we demonstrated that the qualitative results remain robust with a minimal set of moduli, it would be valuable to extend our findings by studying the dynamics through the collective coordinate approximation with additional generalized Derrick modes, potentially improving quantitative accuracy. Another enhancement could involve introducing impurities with one or more optimized parameters. Both of these approaches would increase the complexity of the numerical implementation of the collective coordinates. The collective coordinate analysis could also be extended to kinks with fatter tails ($n > 2$), which could also introduce further technical challenges. Finding an analytical approximation for $A_c(x_0)$ would be another interesting line of work. It would be necessary to generalize perturbative calculations of wobbling kinks \cite{manton1997gradient, forgacs2008negative, barashenkov2009wobbling} to account for the delocalized mode vibrations in multikink configurations. The main challenge in such an analysis is to take into account the force between the kinks.

\section*{Acknowledgements}

JGFC acknowledges financial support from the Brazilian agency FACEPE (Fundação de Amparo a Ciência e Tecnologia do Estado de Pernambuco) grant no. BFP-0013-1.05/23. AM acknowledges financial support from CNPq (Conselho Nacional de Desenvolvimento Científico e Tecnológico), Grant no. 306295/2023-7, and CAPES (Coordenação de Aperfeiçoamento de Pessoal de Nível Superior). Part of the simulations exhibited here were performed in the supercomputer SDumont of the Brazilian agency LNCC (Laboratório Nacional de Computação Científica). The authors would like to thank Andrzej Wereszczynski for fruitful discussions and useful comments on the manuscript.

\end{document}